\documentclass[aps,prl,twocolumn,superscriptaddress,showpacs]{revtex4}
\usepackage{graphicx}
\usepackage{amssymb}

\begin{document}

\title{Electrical switching effect of a single-unit-cell CrO$_2$ layer on rutile TiO$_2$ surface}

\author{Si-Da Li}
\affiliation{Beijing National Laboratory for Condensed Matter
Physics, Institute of Physics, Chinese Academy of Sciences,
Beijing 100190, China}
\affiliation{Department of Physics, Temple University, Philadelphia, Pennsylvania 19122, USA}
\author{Bang-Gui Liu}
\email[Corresponding author:~]{bgliu@iphy.ac.cn}
\affiliation{Beijing National Laboratory for Condensed Matter
Physics, Institute of Physics, Chinese Academy of Sciences,
Beijing 100190, China}
\date{\today}

\begin{abstract}
Rutile CrO2 is the most important half-metallic material with nearly 100\% spin polarization at the Fermi level,
and rutile TiO2 is a wide-gap semiconductor with many applications. Here we show through first-principles investigation that
a single-unit-cell CrO$_2$ layer on rutile TiO$_2$ (001) surface is ferromagnetic and semiconductive with a gap of 0.54 eV, and its electronic state transits abruptly to a typical metallic state when an electrical field is applied. Consequently, this makes an interesting electrical switching effect which may be useful in designing spintronic devices.
\end{abstract}


\maketitle

It is always interesting to search for new spintronic materials and new functional composite materials\cite{wolf,pickett}. Rutile CrO$_2$ is one of the most famous spintronic materials because its nearly 100\% spin polarization at the Fermi level\cite{cro2a,cro2b,cro2c,cro2d,cro2e,cro2f,cro2g,cro2h}, and rutile TiO$_2$ is a semiconductor with a semicondutor gap of 3.0 eV which can be used to make dilute magnetic semiconductors, solar cells and so on\cite{tio2a,tio2b,tio2c}. Recently, there is an ever-large interest to synthesize ultrathin CrO$_2$ films on TiO$_2$ and use them to make interesting devices in order to combine the merits of the CrO$_2$ and TiO$_2$\cite{ctdft,ctexp,ctexp1}. Here, we perform first-principles investigation on epitaxial ultrathin CrO$_2$ layers on rutile TiO$_2$ (001) surface under slab model approach, and thereby find that an epitaxial single-unit-cell CrO$_2$ layer on rutile TiO$_2$ (001) surface is semiconductive, in contrast with metallic state in rutile CrO$_2$ bulk, and its electronic state will transit abruptly to a metallic state when an electric field is applied on it. These results indicate that there is an interesting electric switching effect in this composite material. It could be useful in designing interesting devices. More detailed results will be presented in the following.

We use the plane wave plus pseudo-potential method within the density functional theory (DFT)\cite{dft}, as implemented in VASP package\cite{vasp1,vasp2}, to study the single-unit-cell layer of CrO$_2$ on rutile TiO$_2$ (001) surface. The two Cr-O monolayers are put on the TiO$_2$ (001) surface (infinite number of Ti-O monolayers) and there is the infinite vacuum on the CrO$_2$ layer. We construct finite slab models to describe the infinite system. Our main computational slab model consists of two Cr-O monolayers, seven Ti-O monolayers, and a vacuum layer with a thickness of more than 10 \AA{}, as shown in Fig. 1. A larger slab model with eight Ti-O monolayers is used to make sure that our slab model is reasonable and large enough to capture main physics of the composite system, especially when electric field is applied. The lower TiO$_2$ surface in the slab model is artificial, but it will be shown to have no effect on main electronic properties around the Fermi level. The generalized gradient approximation (GGA)\cite{pbe96} is used for the exchange-correlation functionals, and the semicore Cr p and Ti s states are taken into account. For Monkhorst-Pack grids of k-points, $6\times 6\times 1$ ($4\times 4\times 1$) is used to optimize the structure and $12\times 12\times 1$ ($8\times 8\times 1$) is used to calculate the energies, density of states, and bands for $1\times 1$ ($\sqrt{2}\times \sqrt{2}$) primitive cell. The plane wave cut-off energy is set to 500 eV. The criteria of convergence is 0.01 eV/\AA{} for ionic optimization and $10^{-6}$ eV for electronic self-consistent calculations. The direction of electronic field is along the -z direction.

\begin{figure}[!htbp]
\includegraphics[width=7cm]{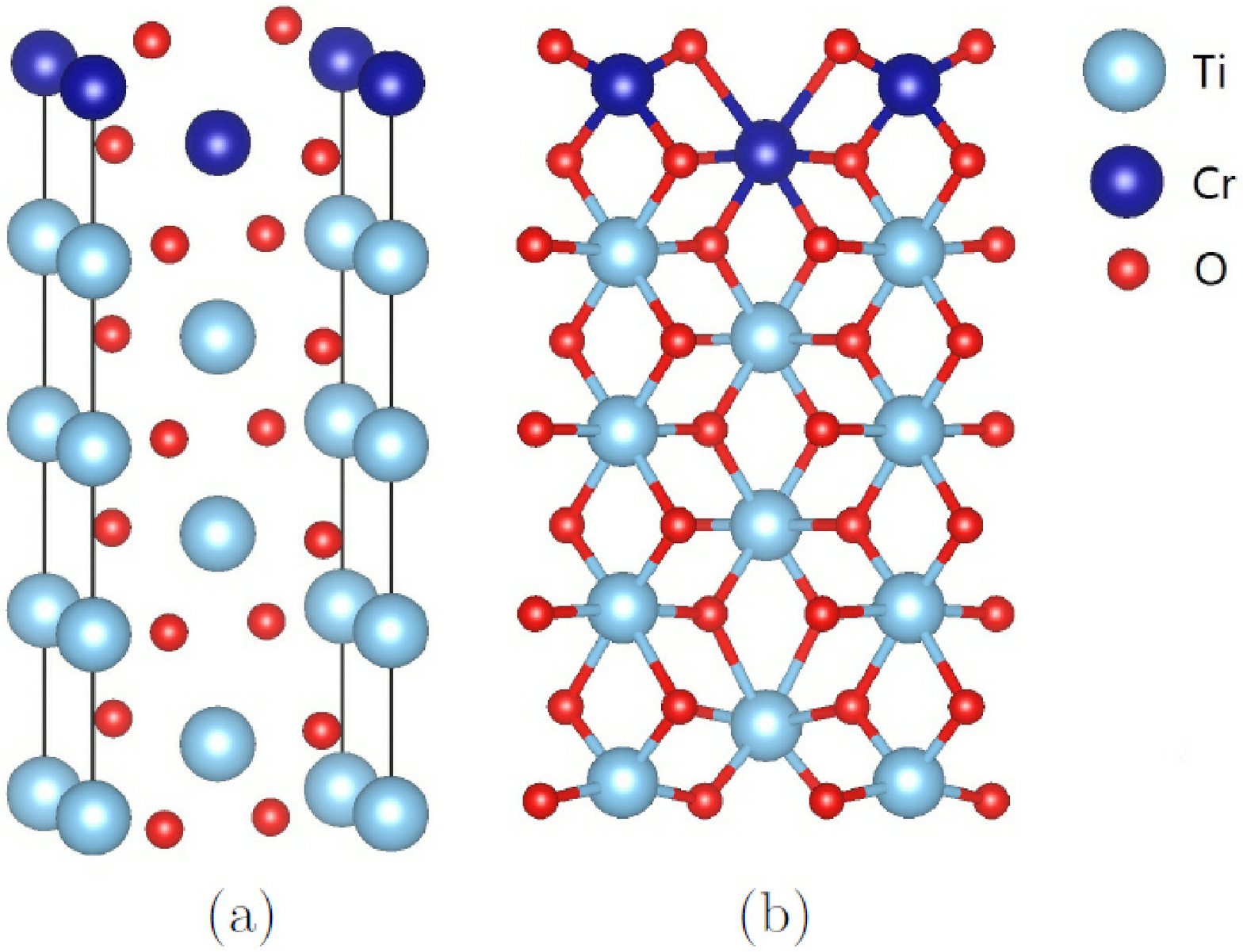}
\caption{(Color online). Crystal structure of the slab model for the single-unit-cell CrO$_2$ layer on the rutile TiO$_2$ (001) surface (a) and that projected on the (010) plane (b).
The lower surface does not affect the main physics.} \label{fig1}
\end{figure}

\begin{figure}[!htbp]
\includegraphics[width=8cm]{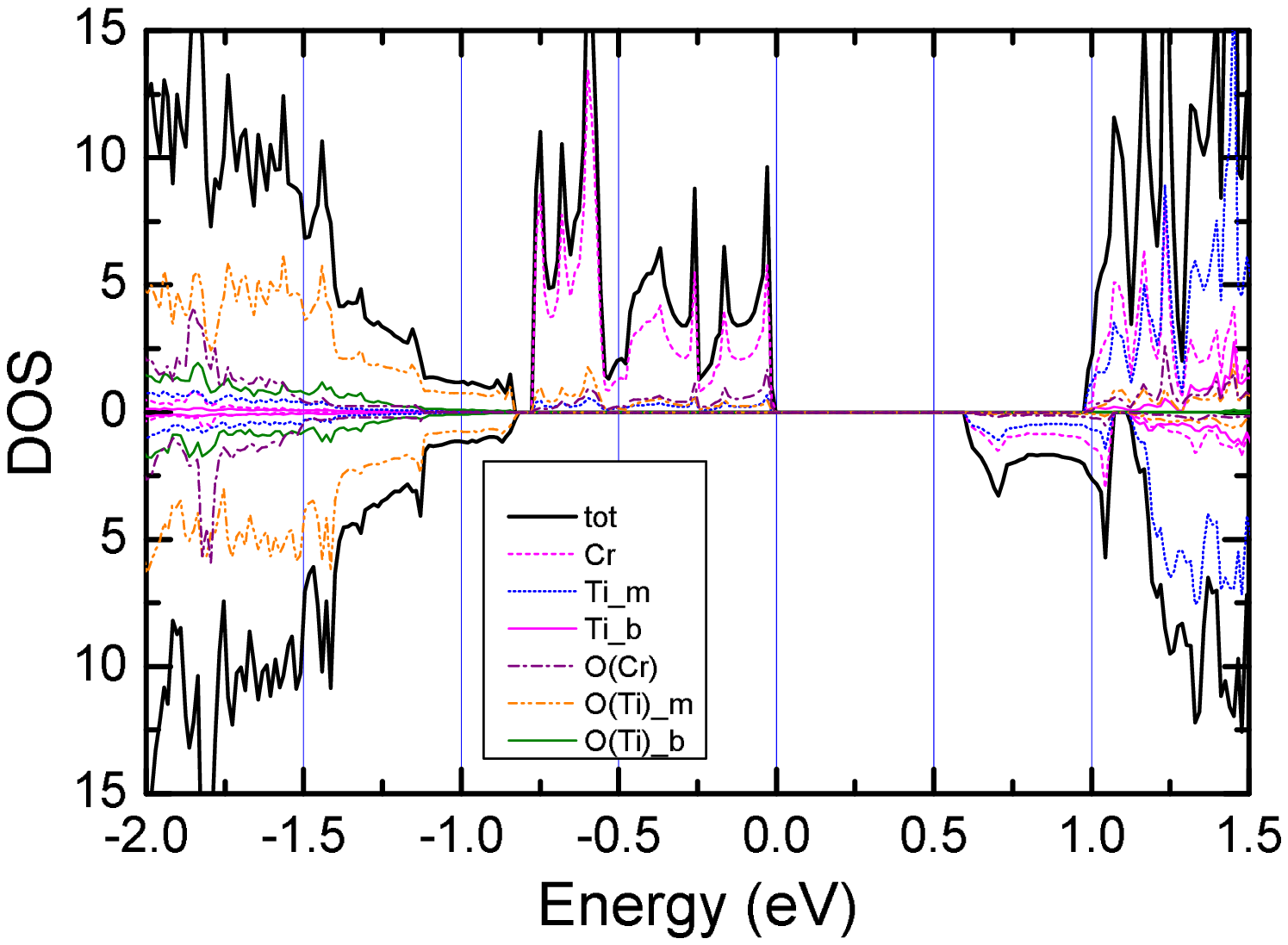}
\caption{(Color online). Total (tot) and atom-resolved density of states (DOS, in state/eV per unit cell) of the ground-state phase without electric field. Cr, Ti\_b, and Ti\_m imply the DOS contributed by Cr atoms, Ti atom in the bottom Ti-O layer, and Ti atoms in the remaining  Ti-O layers together, respectively; and O(Cr), O(Ti)\_b, and O(Ti)\_m are those by O atoms in the corresponding layers. } \label{fig2}
\end{figure}

Rutile TiO$_2$ crystal assumes a tetrahedral structure with lattice constants of 4.594 \AA{} and 2.958 \AA{}\cite{tio2a}, and rutile CrO$_2$ crystal has the same crystal structure with similar lattice constants of 4.421 \AA{} and 2.916 \AA{}\cite{cro2a,cro2b,cro2c}. For the rutile TiO$_2$, our calculated lattice constants are $a=b=4.6503$ \AA{} and $c=2.9718$ \AA{}. They are a little larger than the experimental values because we have used GGA which usually yields a little larger lattice constants for semiconductors. The slab structures have space group Pmm2 (\#25). We divide optimization into two stages. At first stage, we allow only the Cr-O monolayers and the top three Ti-O ones to relax during optimization. With such optimization, we find that the Ti dz$^2$ and dxy orbitals of the bottom Ti-O monolayer, forming the artifact surface, can be near the Fermi level and substantially influenced by applied electronic field. To remove this artificial effect, we make further optimization by allowing the bottom three Ti-O monolayers to relax, fixing the middle Ti-O monolayer only. By doing so, the artificial lower surfaces disappear from electronic states in the energy window of -0.8 and 1.0 eV.

For zero electric field, the ground-state phase is ferromagnetic semiconductor with magnetic moment 4$\mu_B$, because its total energy is lower than the ferrimagnetic structure (magnetic moment 2$\mu_B$) and the antiferromagnetic one (magnetic moment 0$\mu_B$) by 201 and 427 meV (normalized to the atomic number of the ferromagnetic structure), respectively. The ferrimagnetic order is constructed by letting the Cr moments in the two Cr-O monolayers orient anti-parallel, and the antiferromagnetic order by doubling the size of the unit-cell in the xy plane. For the ferromagnetic phase, the most noticeable structural feature is that the surface O atom substantially moves upward so that its bond lengths with the subsurface and surface Cr atoms become 2.432 and 1.669 \AA{}, in contrast with the bulk values 1.918 and 1.913 \AA{}, respectively. As for its magnetic properties, a unit cell contributes a magnetic moment of 4$\mu_B$, most of which comes from the subsurface Cr (2.80$\mu_B$) and the surface Cr (0.91$\mu_B$). We present its total and partial density of states (DOS) in Fig. 2. It is clear that the effect of the lower TiO$_2$ surface is invisible between -1 and 1 eV, and therefore our slab model is reasonable to describe the main electronic and magnetic properties of the single-unit-cell CrO$_2$ layer on rutile TiO$_2$ (001) surface. The DOS between -1.0 and 1.0 eV is contributed by the Cr-O monolayers and the neighboring Ti-O monolayer, with more than 60\% of that between -0.8 and 0 eV originating from the subsurface Cr-O monolayer.

The ground-state phase is still stable when an electric field is applied perpendicular to the surface, until the electric field reaches to $e_c=1.285$ V/\AA{}. At this electric field, there is a phase transition from the semiconducting phase to a metallic one. For convenience, we denote the two phases by A and B in the following. Actually, phase A can be obtained as self-consistent solution of the system for all the applied electric fields between 0 and 1.4 V/\AA{}, and phase B between 1.0 and 2.0 V/\AA{}. Phase A is semiconductive and B metallic, but both are ferromagnetic. Applying electric field does not substantially change the structure and electronic properties for either type-A or type-B phase. We compare their total energies in Fig. 3. Clearly, there is a transition point of electric field, $e_c$, at which the stable phase transits from type A to type B structure. Furthermore, we have estimated the potential barrier of the phase transition from structure A to B by considering those structures which are linear mixture of structures A and B. The barrier of the phase transition at $e_c=1.285$ V/\AA{} is equivalent to 32 meV. This phase transition causes the Cr-O bond lengths to transit from 2.319 and 1.678 \AA{} to 1.921 and 1.774 \AA{}, respectively. The phase transition makes the moments of the surface and subsurface Cr transit from 0.962 and 2.806 $\mu_B$ to 1.652 and 2.349 $\mu_B$, respectively. These big changes over the phase transition should be observed experimentally.

\begin{figure}[!htbp]
\includegraphics[width=8cm]{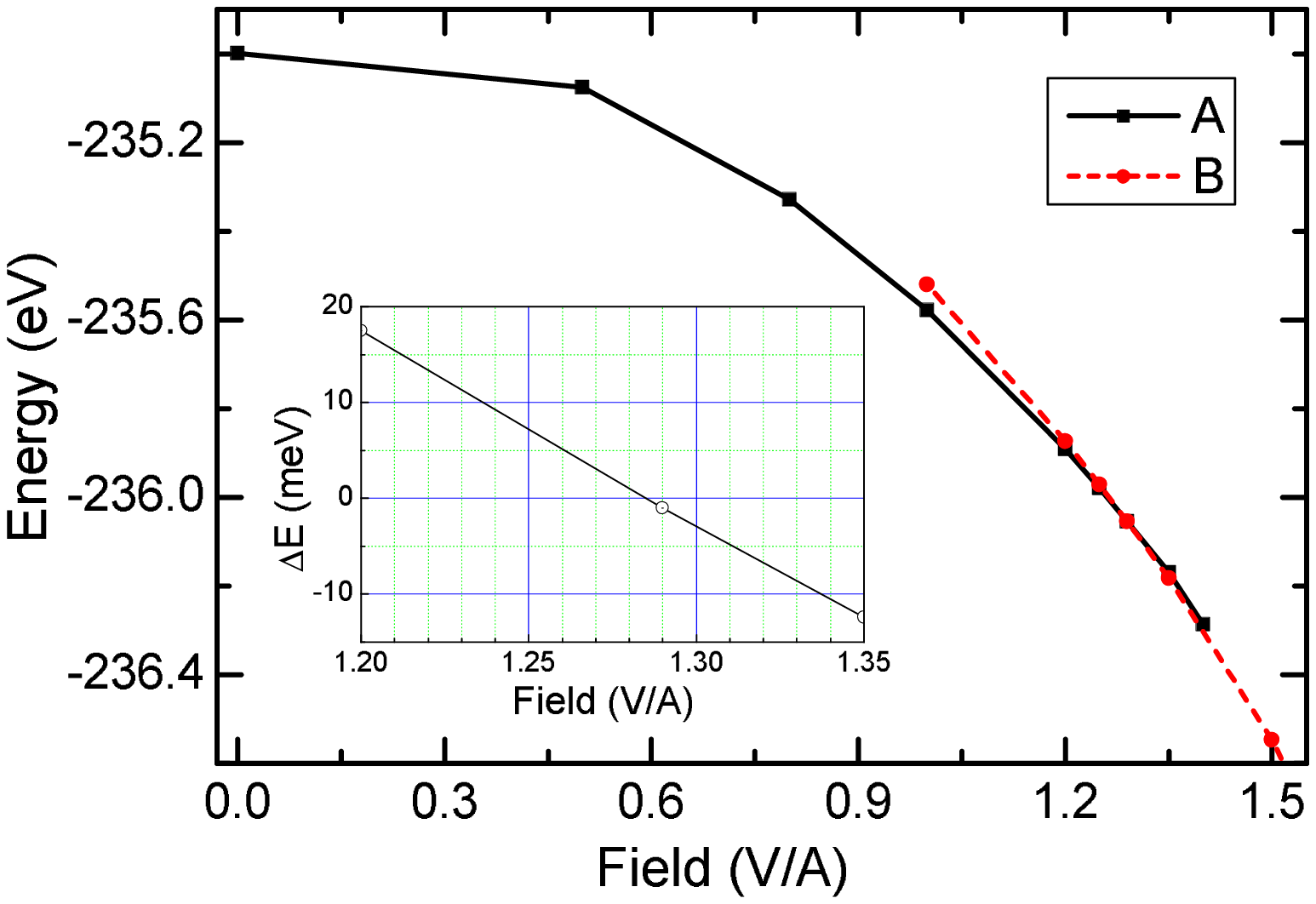}
\caption{(Color online). Total energy comparison between the two ferromagnetic structures: A (solid line) and B (dash line). Structure A is the ground state until the electric field reaches to 1.285 V/\AA, and it is replaced by Structure B when electric field exceeds 1.285 V/\AA. The insert show the total energy difference between B and A, emphasizing the transition point. } \label{fig3}
\end{figure}

\begin{figure}[!htbp]
\includegraphics[width=8cm]{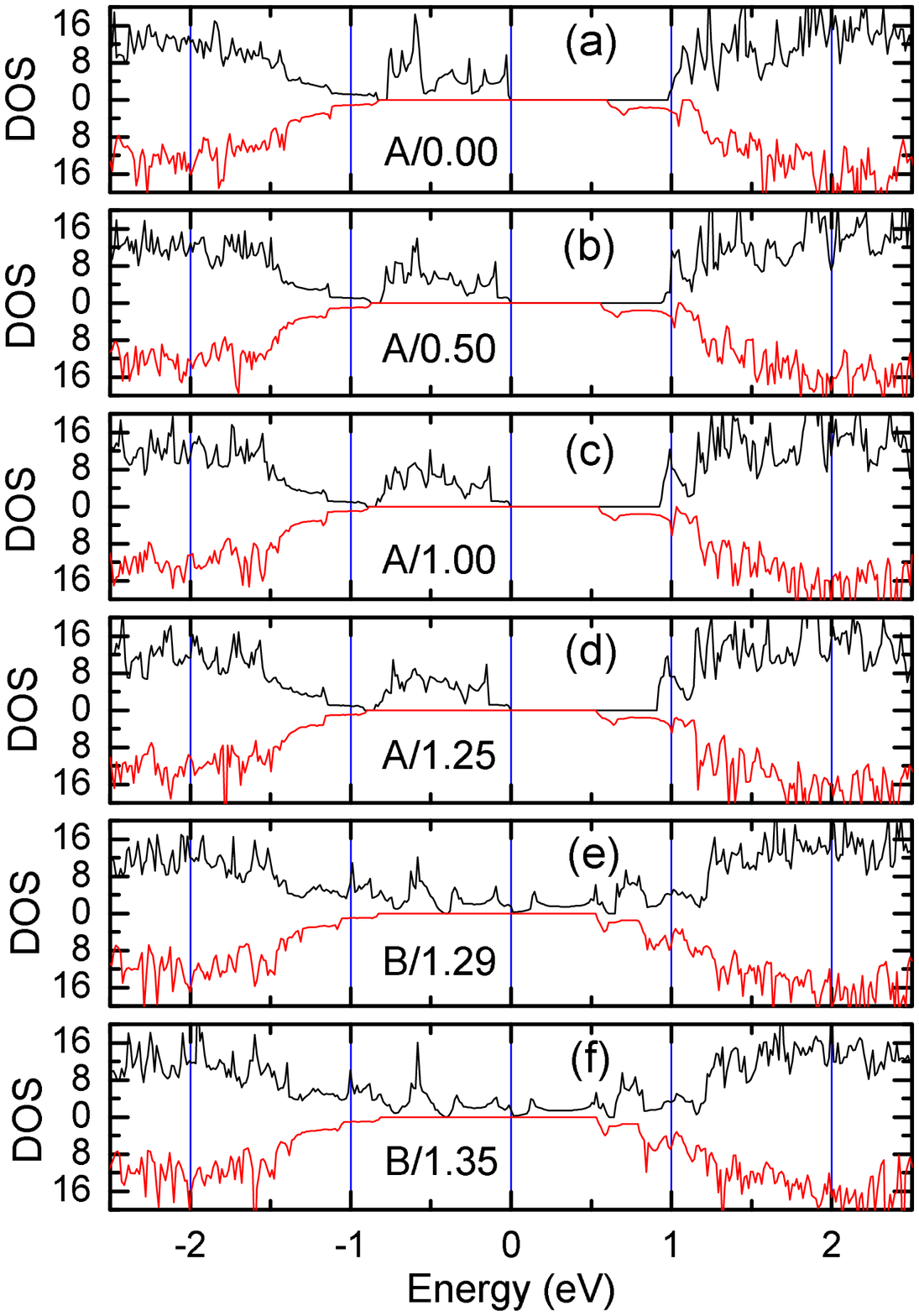}
\caption{(Color online). Spin-resolved density of states (DOS,
in state/eV per formula unit) of the optimized structure for different electric fields: 0.00 V/\AA{} (a), 0.50 V/\AA{} (b), 1.00 V/\AA{} (c), 1.25 V/\AA{} (d), 1.29 V/\AA{} (e), and 1.35 V/\AA{} (f). The lowest-energy structure is type A until the electric field reaches to 1.285V/\AA{}, and it transits to type B at 1.285V/\AA{}. In each panel, the upper part is the majority-spin DOS, and the lower part the minority-spin one.}
\label{fig4}
\end{figure}

In Fig. 4 presented is dependence of the spin-resolved electronic DOS on the applied electric field. It is clear that there is a semiconductor gap of about 0.6 eV and the electronic DOS does not change much when the applied electric field increases from zero to $e_c$, as shown in Fig. 4(a)-(d). When the electric field exceeds $e_c$=1.285V/\AA{}, the semiconducting state is replaced by a metallic one of type B, as shown in Fig. 4(e)-(f). In addition, we can see that the structure B is half-metallic, and the electronic states of structure B around the Fermi level are mostly contributed by Cr and O in the CrO$_2$ layer, and its electronic DOS is not sensitive to the value of electric field. The phase transition is also accompanied by an abrupt transition in the electronic states from a semiconductor of structure A to a metal of structure B. In order to show the abrupt transition, we present in Fig. 5 a layer-resolved density of states of structure B at the transition point $e_c$=1.285V/\AA{} to compare with those of structure A as shown in Fig. 2. At the transition point $e_c$=1.285V/\AA{}, the gap transits from 0.54 eV (phase A) to zero (phase B), and the DOS at the Fermi level from zero (phase A) to 1.3 (phase B) state/eV per unit cell. This electronic state transition from A to B is indeed abrupt, consistent with the abrupt change in the Cr-O bond length over the phase transition.

\begin{figure}[!htbp]
\includegraphics[width=8cm]{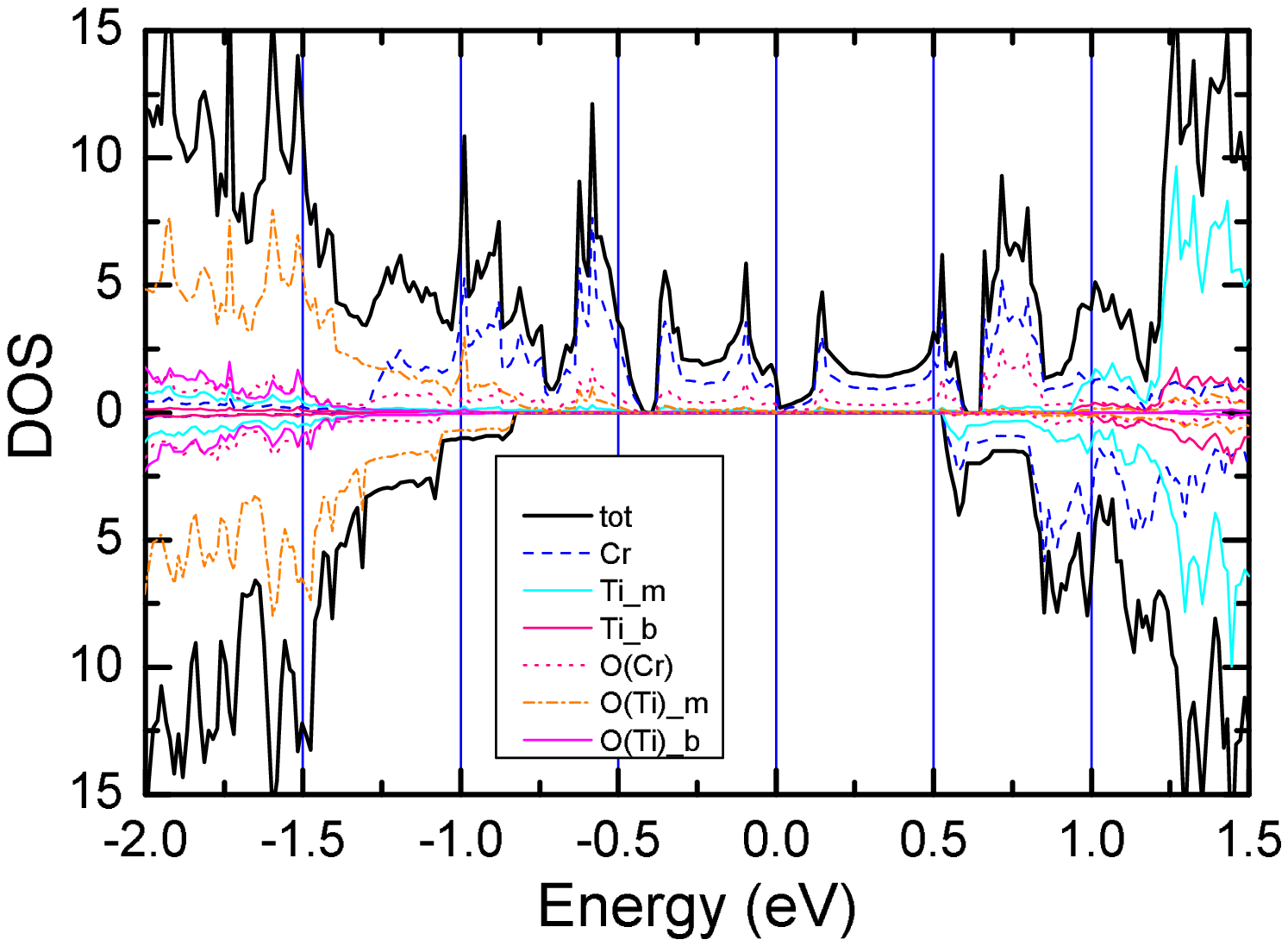}
\caption{(Color online). Total (tot) and atom-resolved density of states (DOS, in state/eV per unit cell) of the B phase with electric field $e_c$=1.285V/\AA{}. Cr, Ti\_b, and Ti\_m imply the DOS contributed by Cr atoms, Ti atom in the bottom Ti-O layer, and Ti atoms in the remaining  Ti-O layers together, respectively; and O(Cr), O(Ti)\_b, and O(Ti)\_m are those by O atoms in the corresponding layers. } \label{fig5}
\end{figure}

We also investigate possible magnetic fluctuations against the ferromagnetic stable state by calculating total energies of the slab model with other magnetic configurations with respect to the stable-state total energy when electric field is applied. Similar ferrimagnetic states and antiferromagnetic states are considered as we did for the field-free case. The ferrimagnetic and antiferromagnetic states are both at least 162 meV higher in total energy than the optimal structures for all the electric fields we calculate.

In addition, we perform total energy calculations by taking the spin-orbit coupling into account. Our results show that the magnetic moment is in the xy plane in both of the two phases. For phase B, the spin-orbit coupling has little effect on the density of states around the Fermi level, causing a reduction of at most 0.05\% in the spin polarization at the Fermi level.

Our structure analysis shows that, for the ground-state phase, there is little overlapping between O atoms, and the surface Cr has four O neighbors and so does the subsurface Cr because it is 2.432\AA{} (1.918\AA{} in the bulk) far from the surface O. This is in contrast with the six O neighbors of Cr in CrO$_2$ bulk. For Cr in this case, its O crystal field appears more like tetrahedral structure than octahedral one in bulk CrO$_2$, and the tetrahedron-like crystal field makes the Cr 3d eg states lower than the t2g ones. The eg-dominated states are fully occupied and the t2g-dominated ones completely empty. As a result, the CrO$_2$ single-unit-cell layer on TiO$_2$ (001) surface, without electric field applied, exhibits a semiconductive behavior. Our magnetic moment and electron number analysis shows that the surface Cr and subsurface Cr have nearly effective chemical valences 5+ and 3+, respectively, in contrast with 4+ in the bulk. When the electric field increases, the surface O atom is compressed downward. At the phase-transition point, the compression makes the bond length between the subsurface Cr and the surface O transit from 2.319\AA{} to 1.921\AA{}, so that the subsurface Cr can be considered to have six O neighbors, being more like that in CrO$_2$ bulk, and hence the metallic phase returns. Despite the configuration of structure B is much more similar with bulk CrO$_2$ than structure A, its conductivity is not contributed by O-O overlapping. The projected DOS of structure B indicate that almost all of those states between -0.4 eV and the Fermi level are contributed by the two Cr-O layers. The subsurface Cr, surface Cr, subsurface O, and surface O contributes 48\%, 13\%, 11\%, and 9\% to the highest occupied states (HOS), respectively. The fact that the subsurface Cr is dominating in HOS and O in the highest Ti-O layer does not affect the HOS also supports our conclusion from bond length analysis. It is the transition of the bond length between the subsurface Cr and the surface O that makes this semiconductor-metal transition. It is interesting that the effective chemical valences of the two Cr ions in the B phase are more similar to that of CrO$_2$ bulk. These make a simple picture for the semiconductive ground-state phase and its field-driven phase transition to the metallic phase.

Now we address experimental feasibility of making a single-unit-cell CrO$_2$ layer on rutile TiO$_2$ (001) surface. Our total energy calculations show that on the TiO$_2$ surface, one single-unit-cell CrO$_2$ layer (two Cr-O monolayers) is much more stable than both one and three Cr-O monolayers. On the other hand, high-quality single-unit-cell FeSe layers have been successfully fabricated on SrTiO$_3$ surface\cite{add}. Therefore, it is reasonable to believe that high-quality single-unit-cell CrO$_2$ layers can be realized on the rutile TiO$_2$ surface.

In summary, we have investigated electronic properties of the CrO$_2$ single-unit-cell layer on TiO$_2$ (001) surface by using a slab approach and density-functional-theory calculations, and thereby found that its ground-state phase is semiconductive and it will transit to a metallic phase when the electric field exceeds a phase-transition point. We elucidate the physical mechanism behind these results. Our further analysis shows that the phase transition is abrupt in the electronic state and therefore there is an electric switching effect in this system. This electric switching effect is interesting and should be useful in designing future spintronics devices.

\begin{acknowledgments}
This work is supported  by Nature Science Foundation of China
(Grant No. 11174359) and by Chinese Department of
Science and Technology (Grant No. 2012CB932302).
\end{acknowledgments}



\begin{thebibliography}{20}

\bibitem{wolf}
S. A. Wolf, D. D. Awschalom, R. A. Buhrman, J. M. Daughton, S. von
Molnar, M. L. Roukes, A. Y. Chtchelkanova, and D. M. Treger,
Science \textbf{294}, 1488 (2001).

\bibitem{pickett}
W. E. Pickett and J. S. Moodera, Phys. Today \textbf{54}, 39
(2001).

\bibitem{cro2a} K. Schwarz,  J. Phys. F \textbf{16},  L211 (1986).

\bibitem{cro2b} K. P. K\"{a}mper et al.,  Phys. Rev. Lett. \textbf{59},  2788 (1987).

\bibitem{cro2c} G. M. Muller, J. Walowski, M. Djordjevic, G.-X. Miao, A. Gupta,
A. V. Ramos, K. Gehrke, V. Moshnyaga, K. Samwer, J. Schmalhorst {\it et al.},
Nat. Mater. \textbf{8}, 56 (2009).

\bibitem{cro2d} Yu. S. Dedkov, M. Fonine, C. Konig, U. Rudiger, G.
Guntherodt, S. Senz, and D. Hesse, Appl. Phys. Lett. \textbf{80},
4181 (2002).


\bibitem{cro2e} J. Kunes, P. Novak, P. M. Oppeneer, C. Konig, M. Fraune,
U. Rudiger, G. Guntherodt, and C. Ambrosch-Draxl, Phys. Rev. B
\textbf{65}, 165105 (2002).


\bibitem{cro2f} R. S. Keizer, S. T. B. Goennenwein, T. M. Klapwijk, G. Miao, G. Xiao, and A.
Gupta, Nature \textbf{439}, 825 (2006).

\bibitem{cro2g} M. S. Anwar, F. Czeschka, M. Hesselberth, M. Porcu, and J. Aarts,
Phys. Rev. B \textbf{82}, 100501(R) (2010).


\bibitem{cro2h} M. S. Anwar and J. Aarts, Phys. Rev. B \textbf{88}, 085123 (2013).


\bibitem{tio2a}  J. K. Burdett, T. Hughbanks, G. J. Miller, J.
W. Richardson Jr, and J. V. Smith,  J. Am. Chem. Soc. \textbf{109}, 3639 (1987).

\bibitem{tio2b} S. Gong and B.-G. Liu, Chin. Phys. B \textbf{21}, 057104 (2012).

\bibitem{tio2c} D. O. Scanlon, C. W. Dunnill, J. Buckeridge, S. A. Shevlin,
A. J. Logsdail, S. M. Woodley, C. R. A. Catlow, M. J. Powell,
R. G. Palgrave, I. P. Parkin {\it et al.}, Nat. Mater. \textbf{12}, 798 (2013).

\bibitem{ctdft} X. Huang, X. H. Yan, Z. H. Zhu, Y. R. Yang, and Y. D. Guo, J. Appl.
Phys. \textbf{109}, 064319 (2011).

\bibitem{ctexp} X. Zhang, X. Zhong, P. B. Visscher, P. R. LeClair, and A. Gupta, Appl. Phys. Lett. \textbf{102\textbf}, 162410 (2013).

\bibitem{ctexp1} K. A. Yates, M. S. Anwar, J. Aarts, O. Conde, M. Eschrig, T. Lofwander, and L. F. Cohen, EPL \textbf{103}, 67005 (2013).


\bibitem{dft} P. Hohenberg and W. Kohn,  Phys. Rev.\textbf{136}, B864 (1964); W.
Kohn and L. J. Sham,  Phys. Rev. \textbf{140},  A1133 (1965).

\bibitem{vasp1} G. Kresse and J. Hafner, Phys. Rev. B \textbf{47}, 558 (1993).

\bibitem{vasp2} G. Kresse and J. Furthmuller, Phys. Rev. B \textbf{54}, 11169
(1996).

\bibitem{pbe96} J. P. Perdew,  K. Burke,  and M. Ernzerhof,  Phys. Rev. Lett.
\textbf{77},  3865 (1996).

\bibitem{add} D. Liu, W. Zhang, D. Mou, J. He, Y.-B. Ou, Q.-Y. Wang, Z. Li, L. Wang, L. Zhao, S. He {\it et al.}, Nat. Commun. \textbf{3}, 931 (2012).

\end{thebibliography}
\end{document}